\documentstyle[12pt]{article}
\begin{document}
\baselineskip = 20pt

\begin{flushright}
{IF-UFRJ/MA1}
\end{flushright}
\vspace*{5mm}

\begin{center}
{\bf The conical singularity method and the energy-momentum tensor near the black hole horizon using the Kruskal coordinates }
\footnote{Work supported in part by Funda\c c\~ao Carlos Chagas Filho-FAPERJ.}
\vspace*{1cm}
\end{center}

\begin{center}
{M. Alves$^{a}$}
\end{center}

\begin{center}
{Instituto de F\'\i sica \\ Universidade Federal do Rio de Janeiro \\

 Caixa Postal 68528  Rio de Janeiro 21970-970 Brazil}

\end{center}

\vskip 1cm

\vspace*{5mm}

\begin{center}
{\bf ABSTRACT}
\end{center}

\vspace*{2mm}

\noindent
We apply the conical singularity method to the two dimensional
version of the Schwarzschild metric 
to obtain the Kruskal coordinates of the black hole in a very simple and direct way. Then we make use of this metric in an approximated version and calculate the expected value of energy-momentum tensor of a massless  quantum field near the horizon, resulting in regular expressions for its components.

\bigskip
\vfill
\par
\bigskip
\bigskip

\par
\bigskip
\bigskip

\noindent $(a)${msalves@if.ufrj.br}

\pagebreak

\vspace*{0.5cm}

{\bf 1- Introduction} 

\bigskip

Relativistic theories of gravitation in two spacetime dimensions have been 
studied intensively for a long time [1,2], the main motivation being the possibility 
of obtaining relevant information on issues of the classical and quantum 
relativistic theory of gravitation in four spacetime dimensions. Works [3] in this 
direction have shown that these two-dimensional models have a rich and 
interesting structure: gravitational collapse and cosmological models  are some of the aspects that can be easily studied within 
this framework. Besides, it has become apparent that these models also have 
remarkable implications in conformal field theories and string-motivated models.

The  two-dimensional gravity model we adopt is given by the  Schwarzchild reduced metric, with the angular parts suppressed. This model has been studied recently [4] as a good description of the motion of a particle near the black hole horizon, where the angular coordinates effects can be neglected.
On the other hand, the Energy-Momentum Tensor (EMT) of a quantum field on this  background  near the horizon shows the same, non physical, singularity as the Schwarzchild coordinates and, if we look for the behavior of these quantities in this region, we must find a description that avoids this singularity. 
It is well known that the metric written in terms of Kruskal coordinates provides us the description without the non physical singularity.   
In this note we apply the method of the conical singularity, already used to compute the Hawking temperature associated to a black hole, to obtain in a simple and direct way the Kruskal coordinates for the Schwarzschild black hole metric. After this, we make use of a few approximations to calculate the components of the semiclassical version of the EMT near the horizon. 

The article is organized as follows. Starting with the two dimensional Schwarzchild metric, we show the basic features of the conical singularity method and  how it is possible to give rise the Kruskal coordinates. The Hawking temperature associated with the black hole is calculated in this meantime. Then we make use of some approximations for the metric in this new form and calculate the EMT components very near the horizon resulting in a regular expression for all the components.

\bigskip

\bigskip

\bigskip

{\bf 2-The conical singularity method }

\bigskip

We start this section with the bidimensional Schwarzschild metric

\begin{equation}
ds^{2}= -(1-{2m\over {r}})\,dt^{2} + (1-{2m\over {r}})^{-1}dr^{2}.
\end{equation}

\noindent There is just one Kretschmann scalar in two dimension spacetime to wit, the scalar curvature that is, to this metric,

\begin{equation}
 R = {4m\over r^3}
\end{equation}

So, by simple inspection, we note that $r=0$ is a real singularity. The $r=2m$ is not itself a singularity since it can be avoided by a sort of coordinate transformation giving us the Kruskal form for the Schwarzschild solution:

\begin{equation} 
ds^{2}= - {16m^{3}\over r}\, e^ {(-{r\over 2m})}\,(dx^{2} - dy^{2})
\end{equation}

\noindent with

\begin{equation}
 x^2 - y^2 = (r -2m)\, e^{  ({r\over 2m })}
\end{equation}

\noindent The conical singularity method (CSM)[5] starts with a Wick rotation of the time coordinate in (1):

\begin{equation}
 t\, \to \,  it
\end{equation}
\noindent so

\begin{equation}
ds^{2}= (1-{2m\over {r}})\,dt^{2} + (1-{2m\over {r}})^{-1}dr^{2}.
\end{equation}

Now we define  two dimensionless variables  $r' = {r\over m} $ and $t' = {t\over m} $ and write (6) as

\begin{equation}
ds^{2}= m^{2}(1-{2\over {r'}})  \biggr\{
dt'^{2} + (1-{2\over {r'}})^{-2}dr'^{2}\biggl\}.
\end{equation}

or
\begin{equation}
ds^{2}= m^{2}(1-{2\over r'})  \biggr\{
dt'^{2} + d(r^{*})^{2}\biggl\}.
\end{equation}

with $r^{*}$ defined by 

\begin{equation}
d(r^{*})^{2}\,=\, \biggl( 1-{2\over r'}\biggr)^{-2} dr'^{2}
\end{equation}

with the solution

\begin{equation}
r^{*}\,=\,r' + 2\ln(\vert r'-2 \vert)
\end{equation}

Let us redefine again both coordinates as
\begin{equation}
\tau\,=\,\alpha t'   
\qquad
\rho\,=\,e^{\alpha r^{*}}
\end{equation}

\noindent that implies
\begin{equation}
dt'^{2}\,=\,{1\over \alpha^{2}}d\tau^{2} 
\qquad
dr^{*}\,=\,{1\over \alpha^{2}\rho^{2}}d\rho^{2}.
\end{equation}

\noindent Using these expressions in (8), we have

\begin{equation}
ds^{2}= \Omega(\rho)\,  \biggr\{
d\rho^{2} + \rho^{2}d\tau^{2}\biggl\} = \Omega(\rho) ds_{flat}^{2}.
\end{equation}

\noindent with 
\begin{equation}
\Omega(\rho)\,=\,m^{2}\biggl(1-{2\over r'}\biggr)\,{1\over \alpha^{2}\rho^{2}}
\end{equation}

\noindent and now the horizon $r'(={r\over m})=2$ is mapped on $\rho = 0$ since, by (11) and (10),

\begin{equation}
\rho\,=\, e^{\alpha r^{*}}\biggl(r'-2\biggl)^{2\alpha}
\end{equation}

We can avoid the singularity at the horizon if we choose in (14)-(15)

\noindent i) the value to $\alpha$ that makes the conformal factor   
 finite  $\Omega(\rho = 0)$  and

\noindent ii) the flat metric $ds_{flat}^{2}$ regular at $\rho=0$.

The expression to the conformal factor is 

\begin{equation}
\Omega(r') \, = \,{e^{-2\alpha r'}\over \alpha^{2} r'} \biggl(r'-2\biggl)^{1-4\alpha} 
\end{equation}

\noindent and the choice $\alpha={1\over 4}$ makes (16) regular at $r'=2$  ($\rho=0$).

To the second choice we note that the  metric 

\begin{equation}
ds^{2}= 
d\rho^{2} + \rho^{2}d\tau^{2}
\end{equation}

\noindent can be a conical or a plane metric. It will be a conical one and therefore with the $\rho=0$ singularity when the angular part has the periodicity less than $2\pi$ or

\begin{equation}
\tau= \alpha t' = \lbrack 0, 2\pi b\rbrack
\end{equation}

with  $ b < 1$. Otherwise, it will describe  a plane  when $b=1$ and,  in this case, (13) will have no problem at $\rho=0$. Our choice is, of course,
$\alpha={1\over 4}$ and  $b=1$. This is the conical singularity method.

As an important consequence of the above result the associated temperature with the system described by (1) can be calculated: since there is no dependence with time coordinate $t$  its periodicity  is related with the temperature $T$ as [5]

\begin{equation}
\Delta t = \Delta (m\,t')= {1\over T}
\end{equation}

and combining (18) and (19), we have

\begin{equation}
\Delta \tau = \Delta(\alpha t')=2\pi
\end{equation}

or

\begin{equation}
m \Delta t' = {1\over T} = {2\pi\over\alpha}
 \end{equation}

With the value of $\alpha = {1\over 4} $ we have the Hawking temperature associated with the black hole described by  (1)
\begin{equation}
T\, =\, {\alpha\over  2\pi m} \,=\, {1\over 8\pi m} \,\,.
 \end{equation}

Besides this well known result, the conical singularity provides us a simple way to write the metric without the apparent singularity: let us just substitute  the value  $\alpha = {1\over 4}$ in (16) and (14) giving

\begin{equation}
ds^{2}= \, {16 m^{2}\over r'} e^{-r'\over 2} \biggr\{
d\rho^{2} + \rho^{2}d\tau^{2}\biggl\}.
\end{equation}

Since the metric (23), modulus the conformal factor, is the flat metric in polar coordinates it can be rewritten as

\begin{equation}
ds^{2}_{flat}\,=\,    d\rho^{2} + \rho^{2}d\tau^{2}  =dx^{2} + dy^{2}
\end{equation}

\noindent with

\begin{equation}
x^{2} +y^{2} = \rho^{2}
\end{equation}

\noindent and, going back to the Minkowski spacetime $ x^{2}\rightarrow -x^{2}$ , we write (23) as

\begin{equation}
ds^{2}= \,  - {16m^{2}\over r'} e^{-r'\over 2} \biggr\{
dx^{2} - dy^{2}\biggl\}.
\end{equation}

\noindent with

\begin{equation}
x^{2} - y^{2} =  e^{{r'\over 2}} \biggr\{r'-2\biggl\}
\end{equation}.

Going back to the original coordinates $t$ and $r$, we have  

\begin{equation}
ds^{2}= \,  - {16m^{3}\over r} e^{-r\over 2m} \biggr\{
dx^{2} - dy^{2}\biggl\}.
\end{equation}

\noindent and

\begin{equation}
x^{2} - y^{2} =  e^{{r\over 2m}} \biggr\{r-2m\biggl\}
\end{equation}.

\noindent exactly as (3) and (4).
 A comment is in order here. The CSM can be understood as a regularization method that provides us a non singular expression to the Schwarzschild geometry with the parameter  $\alpha$  the regularizator  with physical meaning, related with the temperature. The same conclusion remains valid in four dimensions.

\bigskip
\bigskip

{\bf 3- The energy momentum tensor near the horizon} 

In this section we are interested on the calculation of the semiclassical energy momentum tensor (EMT), the fundamental quantity to understand the quantum effects in gravity. In particular, we will obtain the approximate expression for the EMT very near the horizon, useful, for example, to understand the Hawking radiation as a tunneling effect [4,5]. Although regular at the horizon, (28) does not have an explicit dependence with  the new variables, and therefore it can not be used to compute the geometrical quantities only in terms of  $x$ and $y$. As we see below, we can circumvent this difficult by calculating these objects in the $ r\rightarrow 2m $ limit as follow.

Starting with  (28)and (29) as

\begin{equation}
ds^{2}= \,  -{16m^{2} e^{-H}\over H} \biggr\{
dx^{2} - dy^{2}\biggl\}.
\end{equation}

\begin{equation}
x^{2} - y^{2} =  e^{H}(H-1)
\end{equation}

with $H={r\over 2m}$.
\noindent At the horizon  $ r \rightarrow 2m $ or $H\rightarrow 1$
 we have 

\begin{equation}
x^{2} - y^{2} =  e^{H}(H-1)\approx e(H-1)
\end{equation}

\noindent or

\begin{equation}
H\approx {x^{2} -y^{2}\over e} +1
\end{equation}

\noindent The metric (29) or (30) becomes, in terms of the $x$ and $y$ variables

\begin{equation}
ds^{2}= \,  {16m^{2}\over {x^{2}-y^{2}\over e} -1}e^{({x^{2}-y^{2}\over e}-1)} \biggr\{dx^{2} - dy^{2}\biggl\}.
\end{equation}

With the metric in this form, we can calculate the relevant geometrical quantities to wit, the curvature scalar

\begin{equation}
R\,\approx \,{1\over 2m^{2}} \biggr(1+{x^{2}-y^{2}\over e}\biggl)\,\approx {1\over 2m^{2}}
\end{equation}

\noindent and the non zero Christofel symbols

\begin{equation}
\Gamma^{x}_{xx}\, \approx \,-{x\over e}\biggr\{ 2 - {x^{2}-y^{2}\over e}\biggl\}\,\approx -2{x\over e}
\end{equation}

\noindent and

\begin{equation}
\Gamma^{y}_{yy}\, \approx \,{y\over e}\biggr\{ 2 - {x^{2}-y^{2}\over e}\biggl\}\,\approx 2{y\over e}
\end{equation}

\noindent in the limit $ r\rightarrow 2m $.

The  equations we use to obtain the expression of the expected value of the EMT to a massless field , $\langle T_{\mu\nu} \rangle $, follow from the Wald conditions [8]:

\noindent the covariant conservation

\begin{equation}
\nabla_{\nu}\langle T^{\mu\nu}\rangle =0
\end{equation}

\noindent and  its trace that in the first quantum, regularized, correction to massless fields, is proportional to the curvature scalar $R$ 

\begin{equation}
\langle T^{\mu}_{\mu}\rangle = a(h) R
\end{equation}

\noindent where $a(h) $ is proportional to $h$, a purely quantum quantity.

 From (37) and (38) we have the equations

\begin{eqnarray}
\nabla_{\mu}\langle T^{\mu}_{x}\rangle=\partial_{x} \langle T_{x}^{x}\rangle-\partial_{y}\langle T_{x}^{x}\rangle- a\Gamma_{xx}^{x}R+2\Gamma_{xx}^{x}\langle T_{x}^{x}\rangle + 2\Gamma_{yy}^{y}\langle T_{x}^{y}\rangle =0
\nonumber\\
\nabla_{\mu}\langle T^{\mu}_{y}\rangle=\partial_{x}\langle T_{x}^{y}\rangle +\partial_{y}(a R- \langle T_{x}^{x}\rangle  )-a\Gamma_{yy}^{y}R -2\Gamma_{xx}^{x}\langle T_{x}^{y}\rangle=0
\end{eqnarray}.

 We will consider the solutions as near as possible the horizon taking the limit $y \rightarrow x$ in (36)-(37)

\begin{equation}
\Gamma^{x}_{xx}\,\approx -\Gamma^{y}_{yy}\,\approx  -2{x\over e}
\end{equation}

 and we can solve to the EMT components:

\begin{equation}
\langle T_{x}^{y}\rangle\, = \, C_{1} + C_{2} sinh(bx^2) + C_{3} cosh(bx^2)
\end{equation}

\noindent and

\begin{equation}
\langle T_{x}^{x}\rangle\, = \, a R + C_{4} sinh(bx^2) + C_{5} cosh(bx^2)
\end{equation}

\noindent where  $C's$ and $b$ are numerical constants.

 These solutions show that there are  no divergence at the horizon. It is easy to see that inside the horizon ($r < 2m$) we have  the same limit (eq(10) is valid in both regions) as expected when we use the Kruskal coordinates.

\bigskip

\bigskip

\bigskip

{\bf 4- Conclusions} 

In this note we show how to obtain the Kruskal coordinates to the Schwarzschild solution using the conical singularity method in a very straightforward way. The parameter $\alpha$ in (11) can be 
understood as a regularization with physical informations, since the value $ 1\over 4m$ is related with the Hawking temperature that can be obtained by others methods. 

The second result is the approximate expression to the  semiclassical EMT near the horizon. Using the approximate expression to the metric in the Kruskal coordinates the expression to the components of the EMT was found and show us that there is no divergence at the horizon. This result agrees with other authors [6,7].    

\bigskip
\bigskip
\bigskip

 {  {REFERENCES}}

\noindent 1- Jackiw R. 1984 Quantum theory of gravity ed S.Christensen (Bristol: Hilger) p 403.

\noindent 2-Teitelboim C. 1984 Quantum theory of gravity ed S. Cristensen (Bristol: Hilger) 

\noindent\,\,\,\, p. 327.  

\noindent 3-For a complete review with references see Nojiri S. and Odintsov S. 2001  
 
\noindent\,\,\,\, Int.J.Mod.Phys.A 16, 1015.

\noindent 4- R.Jackiw,  Two-dimensional Gravity from Three and Four Dimensions , gr-qc/0310053.

\noindent 5- Amit Ghosh, QED$_{2}$ in Curved Backgrounds, hep-th/9604056.

\noindent \,\,\,\, Dimitri Fursaev and Sergey N. Solodukhin, On the Description of the Riemanniam 

\noindent \,\,\,\, Geometry in the Presence of Conical Defects, hep-th/9501127.

\noindent 6- Balbinot,R., Fagnochi,S., Fabbri,A., Farese,S., Navarro-Salas, J., On the quantum 

\noindent \,\,\,\, stress tensor for extreme 2D Reissner-Nordstrom blach holes,  hep-th/0405263.

\noindent 7- Jingyi Zhang, Zheng Zhao, Phys. Lett. B 618 (2005) 14-22.

\noindent 8- Wald, R.M. Commun.Math.Phys.54, 1 (1977).

\end{document}